\RequirePackage[2020-02-02]{latexrelease}
\documentclass[preprint,aps,amsmath,fleqn]{revtex4}
\usepackage{bm}
\usepackage[dvips]{graphicx}
\usepackage{color}
\usepackage{latexsym}
\usepackage{eufrak}
\usepackage{amsmath}

\newcommand{\eps}{\epsilon}
\newcommand{\Del}{\Delta}

\begin{document}

\title{Semiconductor Laser with Electrically Modulated Frequency}

\author{B. Laikhtman$^{a}$,  G. Belenky$^{b}$ and S. Suchalkin$^{b}$}
\affiliation{a. Racah Institute of Physics, Hebrew University, Jerusalem 91904
Israel \\
b. Department of Electrical Engineering SUNY at Stony Brook, Stony Brook, NY 11794-2350}

\begin{abstract}
 We propose a novel method to control the frequency of semiconductor lasers.  The new technique allows fabricating three-terminal lasers with fast frequency tuning and the possibility to implement intrinsically the linearization of laser frequency sweep. The electrical contact located between the lower undoped cladding and the waveguide together with the upper laser contact enable pumping for optical gain. A voltage applied between the same contact and the contact located under the lower cladding induces space charge limited current (SCLC) across the lower cladding.  Electrons driven into this layer create the space charge. The charge affects the refractive index of the layer and correspondingly the laser frequency. The proposed technique is applicable to any semiconductor lasers. Critical requirements are that free carrier concentration in the lower cladding must be small enough not to affect the SCLC and the laser gain must be high enough to overcome losses introduced by interactivity contact.
 As an example, we present the calculated characteristics of the QCL operated at 10$\mu$m wavelength. Our calculations show that the laser frequency shift can reach GHz range and the  laser tuning speed will be limited by external electronics. Calculations demonstrate that within the range of the selected parameters, the device possesses intrinsically linear relation between the optical frequency and the tuning voltage
\end{abstract}

\maketitle

\section{Introduction}

The current performance of semiconductor lasers, especially quantum cascade lasers (QCLs) are suitable for mid-infrared applications, including light detection, ranging (LiDAR) technologies and free-space optical (FSO) data communication. These applications stand to benefit significantly from the development of mid-infrared lasers with precise frequency control \cite{Haykin,Hayes01,Coldren00,Poulton17,Mitcheo18}. 

Earlier we demonstrated tuning of a single-mode QCL through modulation of cavity refraction index by optical generation of free carriers in the laser waveguide and active area\cite{Suchalkin13}. The physics behind this mechanism is that if the spatial shape of a cavity mode is fixed by the design of the cavity, the variation of the refractive index leads to variation of the mode frequency. In case of optical control of free carrier concentration, the speed of frequency modulation is limited by carrier recombination time and belongs to a MHz range. This limitation can be lifted by application of electrical carrier injection. The effect of electrical
modulation of the complex refractive index in QCL was considered theoretically and experimentally in Ref.\cite{Teissier12}. Electrical frequency modulation of QCL was demonstrated on three terminal devices where control of the electron concentration inside the cavity was achieved by incorporation   of high electron mobility transistor into the laser structure. The change of electron density in the transistor channel leads corresponding cavity index variation \cite{Ohtani14}.

In this paper we propose a novel simple method to control the frequency of semiconductor lasers.  The device frequency control is addressed independently of laser driving voltage. A voltage with frequency tunable over a wide range up to GHz region is applied to non-doped low cladding layer, inducing the  space charge limited current (SCLC)across the layer. The space charge alters the refractive index in the layer which directly affects the frequency of the generated light. The proposed technique is applicable to any semiconductor lasers. Critical requirements are that free carrier concentration in the lower cladding must be small enough not to affect the SCLC and the laser gain must be high enough to overcome losses introduced by interactivity contact. For calculation, as an example, we chose the design of QCL operating at 10$\mu$m wavelength.

The calculation shows that at defined range of the contact doping the dependence of light frequency shift on the applied voltage is practically linear. We emphasize that this linearity is an intrinsic feature of the proposed mechanism and does not require any additional arrangements, such as a complex external circuit. The speed of frequency tuning is constrained by the dynamics of space charge formation and dissipation during voltage modulation. Two primary mechanisms govern these processes: the creation and diffusion (blurring) of the electron cloud due to electron-electron interactions, and the drift of electrons across the cladding layer. Our estimates for the characteristic timescales of both mechanisms fall in the range of 10$^{-12}$ – 10$^{-11}$ seconds. Given these extremely short timescales, it is likely that the tuning speed is primarily limited by the performance of the external electronics.

\section{Structure and calculations}

The proposed laser structure is shown in Fig.1. The active region composed of InAs/InSb superlattice is surrounded by two waveguide layers of equal width. The refractive index of the waveguide layer closely matches the average refractive index of the active region. The system of the active region with waveguide layers is imbedded between cladding layers composed of In$_{0.63}$Al$_{0.37}$As that have a significantly lower refractive index.  The contact layer atop of the upper cladding layer along with another one between the low wave guide and low cladding layer (referred to as the intracavity contact) are for laser pumping. The low cladding layer is mounted on one more contact layer that together with the intracavity contact facilitate the modulation of free electron concentration within the cladding layer.
\begin{figure}
\includegraphics[scale=0.6]{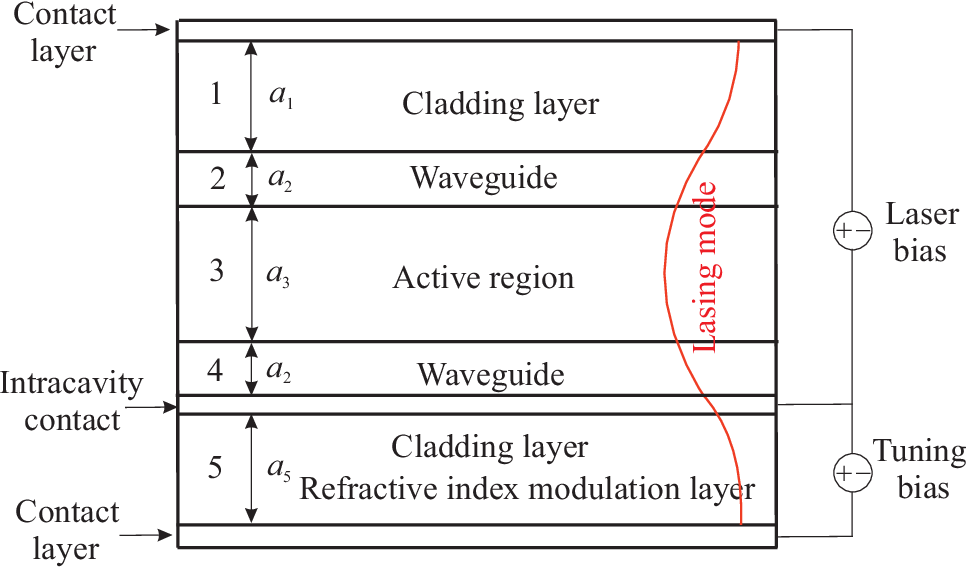}
\caption{\label{fig:1}The structure proposed to frequency modulation of QCL.}
\end{figure}

We considered two possible designs. In the first design, the intracavity contact layer is heavily $n$-doped. Under electric bias applied between the intracavity contact and the lower contact layer, electrons from the intracavity contact are strongly driven into the lower cladding layer, resulting in a SCLC across the cladding.

The second design is similar, but here the lower contact layer is heavily doped, and the direction of the electric bias is reversed compared to the first design. This configuration drives electrons from the lower contact upward into the cladding layer and toward the intracavity contact.

The primary difference between these designs arises from the strongly asymmetric light wave amplitude across the cladding layer. The amplitude decreases from the vicinity of the intracavity contact toward the lower contact. In the first design, where the space charge is concentrated primarily near the intracavity contact, its overlap with the light wave is larger than in the second design, where the space charge is concentrated near the lower contact.

It is possible to expect that in the first design the effect of the refractive index change is stronger than in the second design. On the other hand, stronger overlap leads to stronger losses. 

The calculation that gives a quantitative estimate of the frequency shift dependence on the applied voltage consists of two parts. The first part is the calculation of the light frequency shift resulting from the non-uniform perturbation of the refractive index of the lower cladding layer. The second part is the calculation of the variation of the refractive index in the low cladding layer as a function of applied bias creating SCLC accross the layer.

For calculation of the cavity mode we assume that the dielectric constant of layers 2,3 and 4 is the same, $\eps_{w}$, and dielectric constants of cladding layers $\eps_{1}$  and $\eps_{5}$  are smaller than $\eps_{w}$. Without electron injection to cladding layer 5, i.e. at zero bias between the intracavity contact layer and low contact layer $\eps_{5}=\eps_{1}$. The width of the layers $a_{w}=a_{3}+2a_{2}$, $a_{1}$ and $a_{5}$ are large compared to the width of the contact layers and the last is neglected. If $z$ axis is directed along the wave propagation and $x$ axis is orthogonal to the layers with origin at the intracavity contact then the coordinate dependence of electric field in cavity modes is 
\begin{subequations}
\begin{eqnarray}
&& E_{x} = A_{1}e^{-i\omega t-q_{1}(x-a_{w})}\sin kz  \hspace{2cm} a_{w} < x < a_{w} + a_{1} \ ,
\label{eq:s.1a} \\
&& E_{x} = A_{w}e^{-i\omega t}\sin(k_{w}x + \varphi)\sin kz  \hspace{1cm} 0 < x < a_{w} \ ,
\label{eq:s.1b} \\
&& E_{x} = A_{5}e^{-i\omega t+q_{1}x}\sin kz  \hspace{2.5cm} - a_{5} < x < 0 \ .
\label{eq:s.1c}
\end{eqnarray}
\label{eq:s.1}
\end{subequations}
Here
\begin{equation}
q_{1}^{2} = k^{2} - \eps_{1}k_{0}^{2} \ , \hspace{1cm} k_{w}^{2} = \eps_{w}k_{0}^{2} - k^{2} \ , \hspace{1cm}  k_{0} = \omega/c .
\label{eq:s.2}
\end{equation}
In Eqs.(\ref{eq:s.1a}) and (\ref{eq:s.1c}) we neglected evanescent waves reflected from outer boundaries of the cladding layers (i.e., proportional to $e^{q_{1}x}$ and $e^{-q_{1}x}$ respectively) that makes the calculation significantly simpler. Given $k_{0}$, a mode is characterized by two wave vector components, $k$ in $z$ direction and $k_{w}$ in  $x$ direction. These components are solution to the equations
\begin{subequations}
\begin{eqnarray}
&& kL + \theta = j_{z}\pi \ ,
\label{eq:s.3a} \\
&& k_{w}a_{w} = j_{x}\pi - 2 \arctan\frac{\eps_{1}k_{w}}{\eps_{w}q_{1}} \ ,
\label{eq:s.3b}
\end{eqnarray}
\label{eq:s.3}
\end{subequations} 
where $L$ is the length of the cavity, $\theta$ depends on phases of the mirror reflection coefficients and $j_{x}$ and $j_{z}$ are transverse and longitudinal numbers of the mode.

When electric bias is applied between the intracavity contact layer and low contact layer electrons are injected in cladding layer 5. The injection changes its dielectric constant and it becomes $\eps_{5}=\eps_{1}+\Del\eps_{5}(x)$ where $\Del\eps_{5}(x)$ depends on spatial electron distribution in the cladding layer. $x$ dependence of the dielectric constant leads to more complicated than in Eq.(\ref{eq:s.1c}) $x$ dependence of the electric field in layer 5. This field has to be found from the equation
\begin{equation}
\frac{\partial^{2}E_{x}}{\partial x^{2}} - (q_{1}^{2} - \Del\eps_{5}k_{0}^{2}) E_{x} = 0 \ .
\label{eq:s.4}
\end{equation}
$\Del\eps_{5}(x)$  also modifies boundary conditions for $E_{x}$  at the interface between layers 4 and 5. (Intracavity contact layer between layers 4 and 5 is very thin and its effect can be neglected.) Dielectric constant variation $\Del\eps_{5}(x)$ is small compared to $\eps_{1}$  and it can be considered as a perturbation in both Eq.(\ref{eq:s.4}) and the boundary conditions. As a result, at $-a_{5}<x<0$
\begin{equation}
E_{x} = A_{5}
        \left[
    e^{q_{1}x} - \frac{k_{0}^{2}}{2q_{1}} \ e^{q_{1}x} \int_{0}^{x} \Del\eps_{5}(x')dx' + 
    \frac{k_{0}^{2}}{2q_{1}} \ e^{-q_{1}x} \int_{-a_{5}}^{x} e^{2q_{1}x'} \Del\eps_{5}(x') dx'
        \right]
    e^{-i\omega t} \sin kz \ , 
\label{eq:s.5}
\end{equation}
and Eq.(\ref{eq:s.3b}) is replaced with
\begin{equation}
k_{w}a_{w} = j_{x}\pi - \arctan\frac{\eps_{1}k_{w}}{\eps_{w}q_{1}} - \arctan\left[\frac{\eps_{1}k_{w}}{\eps_{w}q_{1}} \ (1 + \eta)\right]
\label{eq:s.6}
\end{equation}
where
\begin{equation}
\eta = \frac{\Del\eps_{5}(0)}{\eps_{1}} + \frac{k_{0}^{2}}{q_{1}} \int_{-a_{5}}^{0} e^{2q_{1}x}\Del\eps_{5}(x) dx \ .
\label{eq:s.7}
\end{equation}
Wave vector $k$ is determined by the length of the cavity, and perturbation $\eta$ in Eq.(\ref{eq:s.6}) causes a shift of $k_{0}$, such that $k_{0}\rightarrow k_{0}+\Del k_{0}$. This shift induces perturbation of other wave parameters, given by $\Del q_{1}=-(\eps_{1}k_{0}/q_{1})\Del k_{0}$,  $\Del k_{w}=(\eps_{w}k_{0}/k_{w})\Del k_{0}$. As a result, Eq.(\ref{eq:s.6}) leads to the following dependence of the laser frequency shift $\Del k_{0}=\Del\omega/c$ on the perturbation of the refractive index in the low cladding layer
\begin{equation}
        \left[
    \frac{a_{w}}{k_{w}} + \frac{2\eps_{1}k_{w}q_{1}}{\eps_{1}^{2}k_{w}^{2} + \eps_{w}^{2}q_{1}^{2}} \left(\frac{\eps_{w}}{k_{w}^{2}} + \frac{\eps_{1}}{q_{1}^{2}}\right)
        \right] k_{0}\Del k_{0} =
    - \frac{\eps_{1}k_{w}q_{1}}{\eps_{1}^{2}k_{w}^{2} + \eps_{w}^{2}q_{1}^{2}} \ \eta \  .
\label{eq:s.8}
\end{equation}

Application of a voltage between intracavity contact and bottom contact layer induces a current across the lower cladding layer. However, as long as the cladding layer is undoped the current is limited by space charge of electrons that penetrate to the cladding layer from heavily doped intracavity contact. Theory of SCLC was developed by Mott and Gurney\cite{Mott} (see also Refs.\cite{Rose55,Lampert64}). In the first design electron concentration in the cladding layer is
\begin{subequations}
\begin{equation}
n(x) = n_{0} \sqrt{\frac{x_{0}}{x_{0} - x}} \hspace{1cm} -a_{5} < x < 0 \ , 
\label{eq:s.9a}
\end{equation}
where $n_{0}$ is the electron concentration in the intracavity contact layer. In the second design 
\begin{equation}
n(x) = n_{0} \sqrt{\frac{x_{0}}{x_{0} + a_{5} + x}} \hspace{1cm} -a_{5} < x < 0 \ , 
\label{eq:s.9b}
\end{equation}
\label{eq:s.9}
\end{subequations}
Parameter $x_{0}$  depends on the current density $j$,
\begin{equation}
x_{0} = \frac{\eps_{0}j}{8\pi\mu e^{2}n_{0}^{2}}
\label{eq:s.10}
\end{equation}
($\eps_{0}$ is the static dielectric constant and $\mu$ is the electron mobility in the cladding layer. $I-V$ characteristic of the cladding layer is described by the following equation
\begin{equation}
V = \frac{2}{3} \sqrt{\frac{8\pi j}{\eps_{0}\mu}} \left[(a_{5} + x_{0})^{3/2} - x_{0}^{3/2}\right] .
\label{eq:s.11}
\end{equation}

Eqs.(\ref{eq:s.9}) – (\ref{eq:s.11}) are derived by neglecting the diffusion current. The diffusion current is important at small voltage when the difference of the concentration in the doped and undoped regions leads to a potential barrier for electrons at the interface. The width of the barrier is of the order of the Debye screening radius and its height is of the order of temperature assuming that material of both regions is the same. The SCLC takes place at the applied voltage of several tens of volts when the barrier disappears and diffusion is not important. 

The variation of dielectric constant is connected to free electron concentration according to the relation
\begin{equation}
\Del\eps_{5}(x) = - \frac{4\pi e^{2}n(x)}{m\omega^{2}} \ ,
\label{eq:s.12}
\end{equation}
where $m$ is the electron effective mass. Eq.(\ref{eq:s.12}) along with Eqs.(\ref{eq:s.9}) – (\ref{eq:s.11}) define the refractive index in the low cladding layer as a function of coordinate and the applied voltage. 

We made numerical estimates for the waveguide with width of the layers $a_{1}=a_{5}=2\mu$m, $a_{2}=0.15\mu$m, $a_{3}=2\mu$m and refractive indexes $n_{w}=\sqrt{\eps_{2}}=3.5$, $n_{1}=\sqrt{\eps_{1}}=3.3$. Then for wavelength 10$\mu$m  ($k_{0}=0.628\mu$m$^{-1}$, $\omega=1.88\times10^{14}$s$^{-1}$) Eq.(\ref{eq:s.8}) gives for the frequency shift the estimate $\Del\omega/\omega=\Del k_{0}/k_{0}=-C\eta$  with $C=8.9\times10^{-3}$.  

\begin{figure}
\includegraphics[scale=0.7]{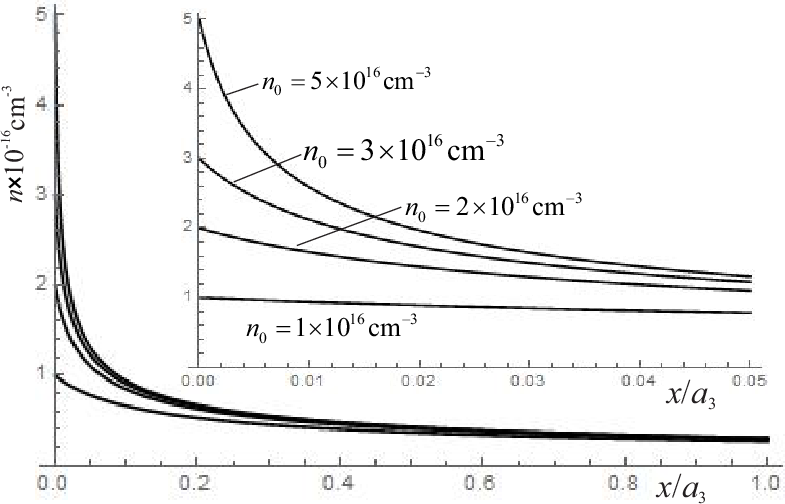}
\caption{\label{fig:n(x)-a}Space charge distribution for voltage 20V and four different contact concentrations. 
The insert shows a part of the main plot close to the contact. Near the contact the electron concentration in the space charge is close to the concentration in the contact. The higher the concentration the faster it falls off with the distance from the contact.}
\end{figure}

The shift of frequency $\Del f=\Del\omega/2\pi$  grows with applied voltage. The details of this growth and their difference for two different designs that we consider hinges on the specifics of space charge distribution that is shown in Fig.\ref{fig:n(x)-a}.
This figure shows the dependence of electron concentration in the cladding layer on the distance from the contact for four different concentrations in the contact at applied voltage 20V. At small distance the electron concentration in the space charge is close to the concentration in the contact. A higher electron concentration leads to stronger screening of the contact, which limits the charge extracting and reduces the distance over which the field can pull these electrons. As a result, the higher the concentration the faster it falls off with the distance. In other words, the higher the concentration the smaller the distance at which the electric field can pull out electrons from the contact.

\begin{figure}
\includegraphics[scale=0.7]{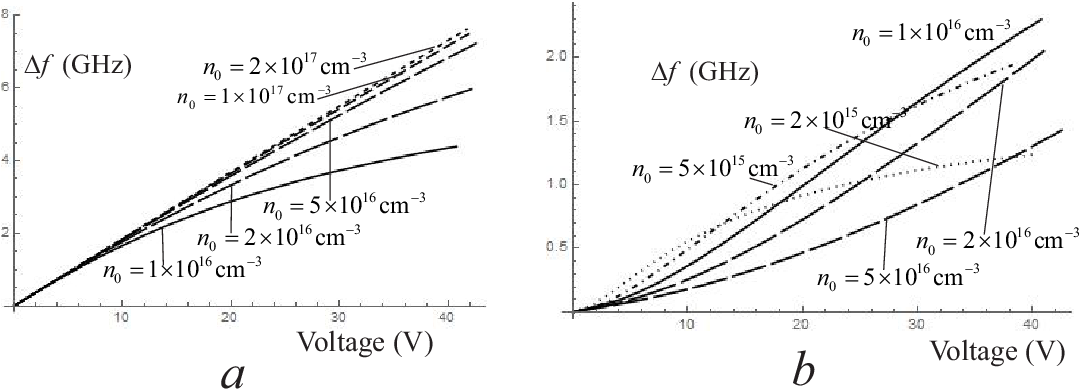}
\caption{\label{fig:Dom(1+2)}Frequency shift $\Del f=\Del \omega/2\pi$ vs voltage between the intracavity contact and low contact layer. $a$ is for the first design and $b$ is for the second design.}
\end{figure}

Frequency shift $\Del f$ as a function of the voltage between the intracavity contact and low contact layer is shown in Fig.\ref{fig:Dom(1+2)} for two different designs: $a$ for strongly doped intracavity contact and electrons pushed from it down to the low cladding layer, and $b$ for strongly doped low contact and electrons pushed up to the low cladding layer. 

When electrons are extracted from the intracavity contact Fig.\ref{fig:Dom(1+2)}$a$, the light wave amplitude decreases with distance from the contact. Consequently, the influence of electrons pulled far from the contact is weaker compared to those closer to it. This phenomenon explains the saturation of $\Del f$ growth with increasing voltage at low electron concentrations.

Notably, at low voltages, the frequency shift increases linearly with the applied voltage. As the electron concentration in the contact increases, the distance over which the voltage can extract electrons decreases, resulting in saturation at higher voltages and an expansion of the linear region. For a doping concentration of $n_{0}=2\times10^{17}$cm$^{-3}$, the maximal deviation from linearity over the voltage range 0–20 V is 0.38\% and in the voltage range 0 - 40V it is 1.2\%.

When electrons are extracted from the lower contact, Fig.\ref{fig:Dom(1+2)}$b$, they first enter a region where the light wave amplitude is very low but increases with distance from the contact. A lower contact concentration allows the electric field to pull the electrons over a greater distance, thereby enhancing their effect. However, if the contact concentration is too small, the refractive index change induced by the electrons becomes very small as well, causing the overall effect to diminish.

In the first case frequency shift $\Del f$ is larger by several times, but for voltage of more than 20V both of them are within the region of a few GHz.

An important question is possible speed of frequency modulation that depends on time necessary for setting in SCLC under variation of the voltage bias. There are two characteristic times relevant to the set in process. One is the Maxwell time characterizing blurring of a electron charge drop, $\tau_{M}=\eps_{0}/4\pi\sigma$ where $\sigma=en_{0}\mu$ is conductivity. The other is an electron drift time across the cladding layer under applied bias, $\tau_{d}=a_{5}^{2}/\mu V$. Electron mobility of III-V compounds at room temperature is  $\gtrsim10^{3}$ cm$^{2}$/V s at doping concentration smaller than $10^{17}$cm$^{-3}$ and varies by 2-3 times depending on composition.\cite{Sotoodeh00} Then $\tau_{M}\lesssim5\times10^{-11}$ s at $n_{0}=10^{16}$cm$^{-3}$ and $\tau_{d}\lesssim10^{-12}$s at $V=10$V. These estimates show that time of setting of SCLC is likely to be limited by an external circuit.

\section{Conclusion}

In this paper, we propose an innovative electrical technique for controlling the frequency of semicoductor lasers. Our method relies on the creation of a space-charge-limited current (SCLC) within the lower laser cladding.

The undoped lower cladding layer is positioned between two contacts: the intracavity contact and the lower laser contact. Applying an electric bias between these contacts induces an SCLC across the lower cladding layer. This space charge modifies the refractive index of the laser cavity, enabling precise frequency control. Importantly, the laser driving current does not pass through the lower cladding, which serves solely to regulate the device's frequency.

This approach is versatile and can be applied to lasers operating at various wavelengths.
 
We calculate frequency shift $\Del f$ as a function of the voltage applied to low cladding for QCL with wavelength 10$\mu$m. In the calculations we neglect the effect of cavity loss on device frequency shift (compare Ref.\cite{Teissier12}).  Our results show that device frequency shift placed in GHz range (see Fig.\ref{fig:Dom(1+2)})

High resolution FMCW LiDAR employs a frequency-swept laser \cite{Karlsson99}.  The light reflected from the target is combined with a local reference signal in a detector, converting the roundtrip distance to a beat frequency.  If the laser frequency sweep is linear, the beat frequency for a stationary target is time independent and can be extracted using a Fourier transform. To the best of our knowledge linear relation between laser frequency and tuning voltage is reached by post-processing of sampled data \cite{Ahn05,Wang16} or external linearization of the laser frequency sweep \cite{Behroozpour17,Martin23}. Both approaches required additional energy and are complicated. Data presented in Fig.3a show 
that relative residual nonlinearity of the $\Del f$ dependence less than 0.38\% can be obtained for QCL with design proposed in this paper.

Thus, three terminal semiconductor laser with controllable space charge limited current across the lower cladding, combines the simplicity of fabrication, fast frequency tuning and possibility to implement   intrinsically the linearization of laser frequency sweep. These futures make the device prospective for fabrication of light detection and FSC systems,  specifically for the development of FMCW LiDARs.

\section{Acknowledgement}

Stony Brook work was supported by ARO contract W911NF2410232


\begin{thebibliography}{9}

\bibitem{Haykin}S. Haykin and M. Moher, \textit{Communication Systems} (John Wiley \& Sons, 2009).

\bibitem{Hayes01}R. R. Hayes, Appl. Opt. {\bf 40}, 6445 (2001). 

\bibitem{Coldren00}L. A. Coldren, “Monolithic tunable diode lasers,” IEEE J. Sel. Topics Quantum Electron., {\bf 6}, 988 (2000).

\bibitem{Poulton17}C. V. Poulton, A. Yaacobi, D. B. Cole, M. J. Byrd, M. Raval, D. Vermeulen, and M. R. Watts, Opt. Lett. {\bf 42}, 4091 (2017). 

\bibitem{Mitcheo18}W. Mitchell, M. S. Hoehler, F. R. Giorgetta, T. Hayden, G. B. Rieker, N. R. Newbury, and E. Baumann, Optica {\bf 5}, 988 (2018). 

\bibitem{Suchalkin13}S. Suchalkin, G. Belenky, T. Hosoda, S. Jung, and M. A. Belkin, Appl. Phys. Lett. {\bf 103}, 041120 (2013).

\bibitem{Teissier12}J. Teissier, S. Laurent, C. Manquest, C. Sirtori, A. Bousseksou, J. R. Coudevylle, R. Colombelli, G. Beaudoin, and I. Sagnes, Optic express {\bf 20}, 1172 (2012).
    
\bibitem{Ohtani14}8. K. Ohtani, M. Beck, and J. Faist, Appl. Phys. Lett. {\bf 104}, 011107 (2014).    

\bibitem{Mott}N. F. Mott and R. W. Gurney, \textit{Electronic Processes in Ionic Crystals}, (Oxford, Clarendon Press 1940).

\bibitem{Rose55}A Rose, Phys. Rev. {\bf 97}, 1538 (1955).

\bibitem{Lampert64}M. A. Lampert, Rep. Prog. Phys. {\bf 27} 329 (1964).

\bibitem{Sotoodeh00}M. Sotoodeh, A. H. Khalid, and A. A. Rezazadeh, J. Appl. Phys. {\bf 87}, 2890 (2000).

\bibitem{Karlsson99}C. J. Karlsson and F. Å. A. Olsson, Appl. Opt. {\bf 38}, 3376–3386 (1999). 

\bibitem{Ahn05}T. J. Ahn, J. Y. Lee, and D. Y. Kim, Appl. Opt. {\bf 44}, 7630–7634 (2005). 

\bibitem{Wang16}Z. Wang, B. Potsaid, L. Chen, C. Doerr, H. C. Lee, T. Nielson, V. Jayaraman, A. E. Cable, E. Swanson, and J. G.  Fujimoto,  Optica {\bf 3}, 1496–1503 (2016). 
    
\bibitem{Behroozpour17}B. Behroozpour, P. A. M. Sandborn, N. Quack, T. Seok, Y. Matsui, M. C. Wu, and B. E. Boser, IEEE J. of Solid-St. Circulation {\bf 52}, 161–172 (2017). 

\bibitem{Martin23}B. Martin, P. Feneyrou, E. Rodriguez, T. Bonazzi, D. Gacemi, M. Berthou, A. Martin, C. Sirtori,  CLEO/Europe-EQEC, 979-8-3503-4599-5 (2023).


\end{thebibliography}
\end{document}